\documentclass[aps,twocolumn]{revtex4}
\usepackage{dcolumn}
\usepackage{bm}
\usepackage{graphicx}
\newcommand{\nc}{\newcommand}
\nc{\bc}{\begin{center}}
\nc{\ec}{\end{center}}

\begin{document}

\title{On  Collective Effects in Cavity Quantum Electrodynamics}
\author  { Bo-Sture K. Skagerstam}
\email{boskag@phys.ntnu.no}
\affiliation
  {\vspace{1mm} \small 
  Department of Physics,  The Norwegian University of Science and Technology,
  N-7491  Trondheim, Norway\footnote{Permanent address} \\
and \\ 
Microtechnology Center at Chalmers MC2, Department of
  Microelectronics 
and Nanoscience, Chalmers University of Technology and 
G\"{o}teborg University, S-412 96, G\"{o}teborg, Sweden }
{\begin{abstract}
\small \noindent We investigate the role of collective effects in  the
micromaser system as used in various studies of the physics of
cavity electrodynamics.
We focus our attention on the effect on large-time correlations
due to multi-atom interactions. The influence of detection efficiencies and collective effects
on the appearance of trapping states at low temperatures is  also
found to be
of particular importance. 
       \\[5mm]
\noindent PACS numbers 32.80.-t, 42.50.-p, 42.50.Ct
\end{abstract}
}

\maketitle


\bc{
\section{INTRODUCTION}
\label{sec:introd}}\ec

The idealized system of a two-level  atom interacting with a second quantized
    single-mode electromagnetic  field, confined in a cavity, plays an important role in the
    study of various  fundamental aspects of quantum mechanics. 
The  micromaser is a remarkable  experimental realization of
    such a simple but fundamental system (for reviews and references see e.g. \cite{Walther88}).  It is therefore also an
 example  of one of the rare systems in Nature
which exhibit a rich structure 
 of physics that can be investigated experimentally
 and which,  at the same time, can be studied by exact theoretical
    methods. 
In the optical regime a microlaser has also been realized experimentally \cite{An94}. Recently this has also been achieved for a one-atom system \cite{Kimble2003}.

Many features of the micromaser system can be regarded to be  of general interest.
Various aspects of
stochastic resonance has e.g. been explored in this system\cite{Buchleitner98}. 
The micromaser also
 illustrates a feature  of non-linear dynamical systems: turning on
randomness may led to an increase of the  signal to noise
ratio \cite{noise94,rek&skag_2000}. 
It can, furthermore,  be argued
that  the micromaser system is a simple
illustration of the conjectured  topological origin of second-order 
phase transitions \cite{Casetti99,rek&skag_2000}.
Trapping states \cite{Filipowicz86,Slosser&Meystre_II} have
been generated in the stationary state of the micromaser system and
therefore  the generation of states with no classical analogue in such
a system is feasible \cite{W&V&H&W,Varcoe_2001}. A basic ingredient of the micromaser is the description of the dynamics in terms of the Dicke model \cite{Dicke54} in the so called rotating wave approximation, i.e. the  of the Jaynes-Cummings (JC) model \cite{Jaynes63}. Early experimental studies involves a confirmation of the JC-model predicted atom revivals \cite{RWK_87}. Recently one has also explicitely demonstrated field mode quantization in a cavity \cite{BSMDHRH_96}. Entanglement of mesoscopic states of the electromagnetic field and an atom in a cavity has also been demonstrated \cite{Haroche2003} in accordance with theoretical considerations (see e.g. Ref. \cite{RSK_2004}). Even though our analysis will focus on the dynamics of the micromaser we observe that the JC model with damping effects included has been realized in ion traps \cite{meekhof&96} and in superconducting systems \cite{nakamura&99}. The coupling of electromagnetic modes to the latter artificial two-level systems has been demonstrated in the laboratory \cite{chir2004,wallraf2004}. We also notice that it has experimentally been verified that phonons can  be confined in semiconductor planar cavities \cite{trigo2002}. It is therefore not unlikely that the present analysis may find applications in systems similar to the micromaser system but realized in a completely different physical framework.

The paper is organized as follows. In Section \ref{sec:dynamics} we outline the dynamics of a typical experimental setup of the two-level system interacting with a single-mode of the radiation field. Long-time correlations are discussed in Section \ref{sec:longcorr} and corrections due to detection efficiencies are discussed in Section \ref{sec:detection}. Collective effects due to the finite probability of having two atoms at a time in the cavity are analyzed in Section \ref{sec:collective} together with possible effects on the detection of trapping states. In Section \ref{sec:final} we summarize our work and indicate effects on the phase transitions of the micromaser system in the so called large-$N$ limit.
 
\vspace{1cm}
\bc{
\section{MICROMASER DYNAMICS}
\label{sec:dynamics}}\ec
 
   In our analysis we consider the following  typical realization of
   the micromaser. The pump atoms which enter the cavity 
    are at resonance with the radiation field of the cavity, and are also assumed to be prepared in the excited state. The injection intervals between the incoming atoms
   are assumed to be Poisson-distributed.
In terms of the dimensionless atomic flux parameter $N=R/\gamma$, where $R$ is the rate 
injected atoms and $\gamma$  is the damping rate of the cavity, 
the stationary photon number probability distribution  is then described by a diagonal density matrix with diagonal elements which are
well known  \cite{Filipowicz86} and  are given by
\begin{equation} \label{p_n_eksakt}
  {\bar p}_n = {\bar p}_0 \prod_{m=1}^{n} 
  \frac{n_b \, m +  N q_m}{(1+n_b) \, m }~~~.
\end{equation}
\noindent
   Here $q_m \equiv q(x)=\sin^2\left(\theta
           \sqrt{ x }\right)$, with $x=m/N$, and
   where we have defined the natural dimensionless  pump parameter $\theta = g\tau \sqrt{N}$
   in terms of the atomic transit time $\tau$. 
   Furthermore,  $g$ is the single photon Rabi frequency at zero detuning
   of the JC-model \cite{Jaynes63}.
   The overall constant ${\bar p}_0$ is determined by $\sum _{n=0}^{\infty}{\bar p}_n =1$.

   The theory as  developed in Refs.\cite{Filipowicz86,Guzman89}
   suggests the existence of various phase transitions
   in the large-$N$ limit as the parameter $\theta$
   is increased.   A natural order parameter
   is then the average photon ``density'' $\langle x\rangle $, 
   where $\langle ~\rangle$ denotes an average  with
   respect to the distribution Eq.~(\ref{p_n_eksakt}). 
   An exact large-$N$ limit treatment of the micromaser
   phases structure and the corresponding critical fluctuations in
   terms of a conventional correlation length  was presented in
   Refs.\cite{ElmforsLS95}. Spontaneous jumps in  $\langle n\rangle /N$ 
   and  large correlation lengths close to micromaser phase transitions have
   actually been observed experimentally \cite{Walther88,Walther97}.
    Several new
   intriguing physical properties of the micromaser system are
   unfolded when the theoretical 
analysis is extended to a more
 general setup of the parameters available in the micromaser
 system than those considered here \cite{Rekdal&Skagerstam&99a}.

\vspace{0.5cm}
\bc{
\section{LONG-TIME CORRELATION EFFECTS}
\label{sec:longcorr}
}\ec

Let us now consider long-time correlations in the large-$N$ limit
   as was first introduced in Refs.\cite{ElmforsLS95}.
   These correlations are most
   conveniently expressed in terms of the continuous-time
   formulation of the micromaser system \cite{Lugiato87}. 
   The vector $p$ formed by the diagonal density matrix elements of the photon
   field then obeys the differential equation 
\begin{equation} \label{evolution_1}
 \frac{dp}{dt}= - \gamma Lp ~~~,
\end{equation}
where
   $L=L_C -N(M-1)$. Here $L_C$ describes the damping of the cavity , i.e.

\begin{eqnarray}
 && ( L_C)_{nm} = (n_b+1)[ \, n \delta_{n,m} - (n+1) \delta_{n+1,m} \, ] 
  \nonumber \\ && 
         ~~~~~~~~~~~   +  n_b[ \, (n+1)\delta_{n,m}
              - n \delta_{n,m+1} \, ] ~~,
\end{eqnarray} 

   \noindent
   and $M=M(+) +M(-)$, where $M(+)_{nm} = 
   (1-q_{n+1}) \delta_{n,m}$ and $M(-)_{nm} = q_{n} \delta_{n,m+1}$
   have their origin in the 
JC-model \cite{Jaynes63,ElmforsLS95}.
   The lowest eigenvalue $\lambda_0=0$ of $L$ then determines the
   stationary equilibrium solution ${\bar p}=p^{(0)}$ as given by
   Eq.~(\ref{p_n_eksakt}).
${\cal P}_{s}(\tau)= \mbox{Tr}[M(+){\bar \rho}]= {\bar u}^{T}M(s){\bar p}$ then is the probability that an atom is found in
the  state $s=\pm$, 
where $+(-)$ denotes the excited(ground) state, after it leaves the
microcavity. The vector ${\bar u}^T$ is the transpose of the vector ${\bar u}$  with 
all entries equal to 1.  Here we have used the fact that the vector ${\bar u}^{T}$ simply represents the trace operation and ${\bar \rho}$ is the diagonal density matrix of the cavity field. 
When the injection intervals between the incoming atoms are Poisson-distributed, as we have assumed is the case, the joint probability, ${\cal P}(s_1,s_2,t)$, of observing two atoms
in the states $s_1$ and $s_2$ with a large time delay $t$ between them  can be written  in the form
\begin{eqnarray}
 \label{joint_t} 
  {\cal P}(s_1,s_2,t) & = & \mbox{Tr}[S(s_2) \,   
  e^{-\gamma L t} \, S(s_1){\bar \rho}] 
 \nonumber \\  
  &=&  
  {\bar u}^{T} M(s_2) \,   
  e^{-\gamma L t} \, S(s_1)~ {\bar p}   ~~. 
\end{eqnarray}
The time delay $t$
corresponds to a large number, $k \simeq Rt$, of unobserved atoms between the two detections. In Eq.(\ref{joint_t}) we make use of the propagation matrix
\begin{equation} \label{soperator}
    S(s) = (1+L_{C}/N)^{-1}M(s)~~~,
\end{equation}
where $S= S(+)+S(-)$. $S$ is a so called stochastic matrix (see e.g. Refs.\cite{fitz79,reichl98}) with left ($u^{(n)}_S$) and right ($p^{(n)}_S$) eigenvectors corresponding to the eigenvalue
 $\kappa_n$, where $n=0,1,2,...$, such that $1=\kappa_0 > \kappa_1 > ... \geq 0$. The stationary distribution as given by  Eq.(\ref{p_n_eksakt}) corresponds to the eigenvalue $\kappa_0 =1$, i.e. $u^{(0)}_S = {\bar u}$ and $p^{(0)}_S={\bar p}$. The spectral decomposition of $S^k$, where $k=0,1,...$, i.e.
\begin{equation} \label{eq:spectral}
S^k = \sum_{n=0}^{\infty}\kappa^k_{n}p_S^{(n)}u_S^{(n)T} ~~~,
\end{equation}
is useful in many of the numerical calculations presented below. Here the normalization is such that $u_S^{(n)T}p_S^{(m)} = \delta_{nm}$.
The joint probability  ${\cal P}(s_1,s_2,t)$ is symmetric, i.e. ${\cal P}(s_1,s_2,t)= {\cal P}(s_2,s_1,t)$, and is  properly normalized, i.e.
$\sum_{s_1 s_2} {\cal P}(s_1,s_2,t) =1$ since $S{\bar p}= {\bar p}$ and ${\bar u}^T{\bar p}=1$. In the
 original discrete
   formulation of the micromaser system \cite{Filipowicz86} the corresponding
joint probability ${\cal P}_k(s_{1},s_{2})$ of
 observing two atoms
in the states $s_1$ and $s_2,$ with $k$ unobserved atoms between, 
   can be written  in the form
\begin{equation} \label{joint_k}
    {\cal P}_k(s_{1},s_{2}) = \mbox{Tr}[S(s_2)S^{k}S(s_1){\bar \rho}]= {\bar u}^{T}M(s_2)S^{k}S(s_1){\bar p}~~~.
\end{equation}
This  joint probability is also properly normalized, i.e.
$\sum_{s_1 s_2} {\cal P}_k(s_{1},s_{2}) =1$.
For a sufficiently large $k$  
and with $k\simeq Rt$ it now follows that
\begin{equation}
 \label{cal_P} 
  {\cal P}_k(s_1,s_2) =  {\bar u}^{T} S(s_2) \,   
  e^{-\gamma L t} \, S(s_1)~ {\bar p}  
    = {\cal P}(s_1,s_2,t) 
\end{equation}
in the large-$N$ limit \cite{ElmforsLS95} and where  $L$ is as in Eq.~(\ref{evolution_1}). 
A formal way to see the validity of Eq.(\ref{cal_P}) is to notice that $(1+L_{C}/N)^{-1}= 1-L_{C}/N +{\cal O}(1/N^2)$. We can then write
\begin{equation} \label{eq:expansion}
    S = 1- \frac{1}{N}(L_{C}-N(M-1))~~~,
\end{equation}
apart from  terms of order $L_C(M-1)/N$ and higher orders in $1/N$. Below we will find expansions of the form Eq.(\ref{eq:expansion}) very useful in the analysis of detection efficiencies.
We observe that the joint probabilities ${\cal P}_k(s_1,s_2)$ are symmetric in $s_1$ and $s_2$ \cite{ElmforsLS95} as in the continuous-time formulation, i.e. the order in which the atoms are measured is irrelevant. 
Other definitions for such  joint probabilities have appeared in the 
literature. In the definition for the two-time coincidence
probability used in e.g. 
Refs.\cite{Herzog94,Englert98}, $S(s_1)$ in Eqs. (\ref{joint_t}) and (\ref{joint_k}) 
 is replaced by $M(s_1)$. With the Poisson-statistics assumption behind the derivation  
of these equations this would, however, seem like an unnatural thing to
do. Since the correlation length to be defined and used below is only sensitive to the eigenvalues of the operator $L$, our discussions below will not, in the end, be effected by such a modification, at least not in the large-$N$ limit. With detector efficiencies $\eta_+$ and $\eta_-$ for detecting atoms in the excited or ground state respectively, the sequence probabilities defined and used in Ref.\cite{Johnson2001} are identical to our joint probabilities Eq.(\ref{joint_k}) in the discrete formulation of the micromaser if $\eta_+=\eta_- = 100\%$. For $k=0$ the joint probability ${\cal P}_k(s_{1},s_{2})$ reduces 
to ${\cal P}_{s_{1}s_{2}}(\tau)$, i.e. the probability that the
next atom is in the 
state $s_{2}=\pm$ if the previous atom  has been found in the 
state $s_{1}$. ${\cal P}_{s_{1}}(\tau)$ exhibits the  experimentally 
observed revivals in microcavity systems \cite{RWK_87,BSMDHRH_96} and
in e.g. 
ion-traps \cite{meekhof&96}. ${\cal P}_{s_{1}s_{2}}(\tau)$ exhibits in
addition 
so called pre-revivals \cite{ElmforsLS95}. 

\begin{figure}[htp]
\unitlength=0.5mm
\begin{picture}(160,140)(0,0)
\includegraphics{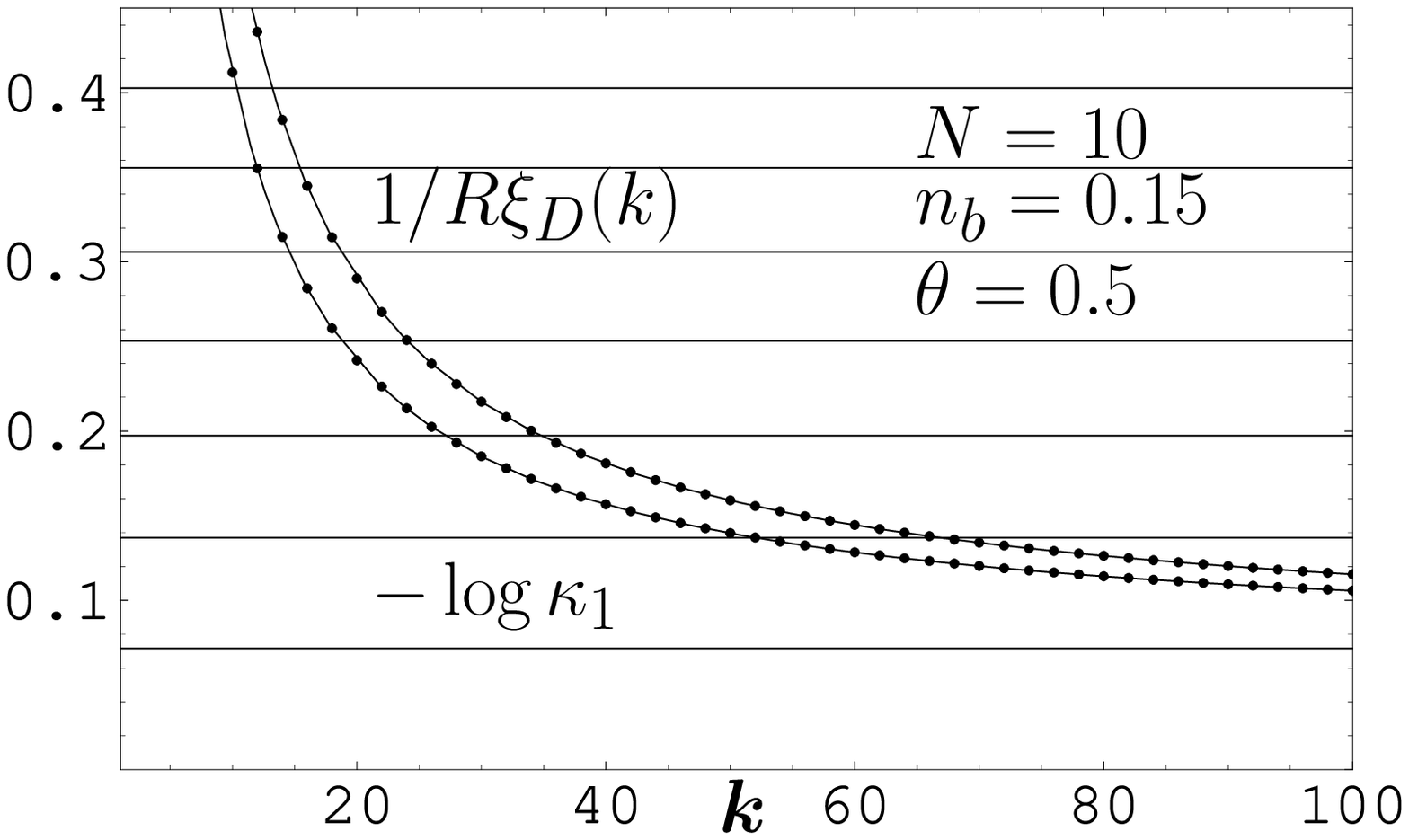}
 \end{picture}
\caption[]{\protect\small 
           The convergence of the correlation length  $1/R
           \xi_{D}(k) = -\log(|\gamma_{D}(k)|)/k$ as a function 
           of the number $k\simeq Rt$ of unobserved atoms leaving the cavity
           for $\theta =0.5$, $N=10$ and $n_b =0.15$ in the case of
           the discrete formulation of the micromaser system with an analytical fit using only
           the next to the leading eigenvalue $\kappa_1$ of the stochastic matrix $S$ as
           defined in the main text. The corresponding correlation length is $R\xi \approx 14$. The upper curve
           corresponds to the definition of joint probabilities of the present paper. The lower curve
           corresponds to a redefinition $S(s_1) \rightarrow M(s_1)$ in the joint probabilities Eq.(\ref{joint_k}). 
The horizontal lines corresponds $-\log(\kappa_n)$ for $n=1,...,7$.}
\label{conv_fig}
\end{figure}

\begin{figure}[htp]
\unitlength=0.5mm
\vspace{5mm}
\begin{picture}(160,140)(0,0)
\includegraphics{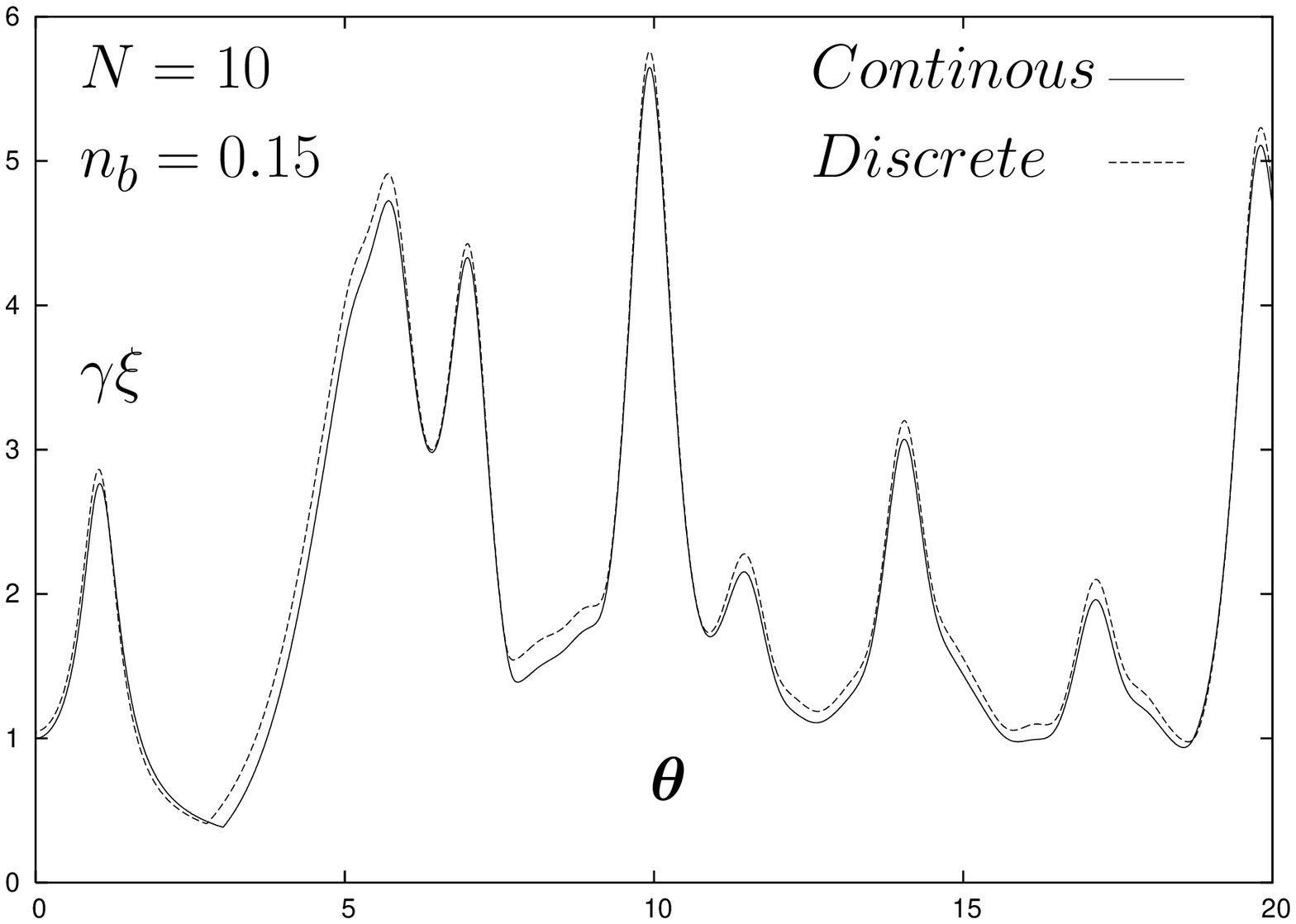}
 \end{picture}
\vspace{5mm}
\caption[]{\protect\small 
           The correlation length  $\gamma \xi$ as a function of 
           $\theta$ for $N=10$ and $n_b=0.15$ in the case of the continuous-time (solid
           line)
and the discrete (dashed line) formulation of the micromaser system as
described in the main text. }
\label{compare_fig}
\end{figure}

 A properly normalized correlation function  $\gamma_C(t)$ can now be defined  and
   expressed in different but equivalent manners, i.e. 
\begin{eqnarray} \label{def:gamma_C} 
  \gamma_C(t)\equiv \frac{ \langle ss \rangle _t - \langle s \rangle ^2} 
                     {  1 - \langle s \rangle ^2 } &=& 
 \frac{ {\cal P}(+,+,t) - {\cal P}(+)^2}{ {\cal P}(+) {\cal P}(-)} \nonumber \\ 
= \frac{ {\cal P}(-,-,t) - {\cal P}(-)^2}{ {\cal P}(+) {\cal P}(-)} &=& 
\frac{ {\cal P}(+){\cal P}(-) - {\cal P}(+,-,t)}{ {\cal P}(+) {\cal P}(-)} 
~,\nonumber \\
\end{eqnarray} 
   where $\langle ss \rangle_t = \sum _{s_1,s_2} s_1 s_2 {\cal P}(s_1,s_2,t)$ 
   and $\langle s \rangle = \sum _s s \, {\cal P}(s)$. This correlation 
   function satisfies $-1 \leq \gamma_C(t) \leq 1$.  At large times  
   $t\rightarrow \infty$, we then define the atomic beam correlation length  
   $\xi_C$ by \cite{ElmforsLS95} 
\begin{equation} \label{gamma_C} 
   \gamma_C(t) = \gamma_Ce^{-t/\xi_C}~~~, 
\end{equation}
   which then  is determined by the next-to-lowest eigenvalue $\lambda_1$ of $L$, i.e. $\gamma\xi_C=1/\lambda_1$.
In the discrete formulation of the micromaser system we make the replacement ${\cal P}(s_{1},s_{2},t) \rightarrow {\cal P}_k(s_{1},s_{2})$ in Eq.(\ref{def:gamma_C}) and, for a sufficiently large $k$, we then define, in a similar manner, the correlation length  $\xi_D$  by the expression
\begin{equation} \label{gamma_D} 
   \gamma_D(k) = \gamma_De^{-k/R\xi_D}~~~.
\end{equation}
 In the large-$N$ limit one can show  that $\xi_C$ and $\xi_D$ converge to the same limit and we therefore write $\xi \equiv \xi_C=\xi_D$ for sufficiently large $N$ \cite{ElmforsLS95}. For photons a similar analysis leads us to  a correlation length $\xi_{\gamma}$. It
   follows that the correlation lengths so defined are identical in the large-$N$ limit, i.e. 
   $\xi_{\gamma}=\xi $ \cite{ElmforsLS95}.
In Fig. \ref{conv_fig} we illustrated the convergence of  $1/R\xi_{D}(k) \equiv -\log(|\gamma_{D}(k)|)/k$ to its asymptotic value $1/R\xi = -\log(\kappa_1)$ for a typical experimental setup of the micromaser system. We observe that spectral resolution Eq.(\ref{eq:spectral}) and the definition Eq.(\ref{def:gamma_C})  modified to yield $ \gamma_D(k)$ leads to
\begin{equation} \label{eq:cns}
   \gamma_D(k) =\sum_{n=1}^{\infty}c_n\exp(-k\log(1/\kappa_n))~~~, 
\end{equation}
where we have defined
\begin{eqnarray} 
   c_n  &=& 
\sum_{s_1s_2}\frac{{\bar u }^T M(s_1)p^{(n)}_{S}u^{(n)T}_{S}S(s_2){\bar p}}{1 - \langle s \rangle ^2} \nonumber \\ &=& \frac{{\bar u }^T M(+)p^{(n)}_{S}u^{(n)T}_{S}S(+){\bar p}}{{\cal P}_{+}{\cal P}_{-}}~~~. 
\end{eqnarray}
The analytical fits in Fig.\ref{conv_fig}, $1/R\xi_D=c_1/k - \log(\kappa_1)$,  are based on using only the $n=1$ term in Eq.(\ref{eq:cns}). We have found that this approximation works for all $\theta$ despite the fact that $N$ is small in this case.  The numerical value of $c_1 $ can easily be computed numerically given the micromaser parameters. The upper curve in Fig.\ref{conv_fig} corresponds the definition Eq.(\ref{joint_k}) of joint probabilities and leads to $c_1 \approx 3.40$. The lower curve in Fig. \ref{conv_fig}
           corresponds to a redefinition $S(s_1) \rightarrow M(s_1)$ in the joint probabilities Eq.(\ref{joint_k}) and leads to $c_1 \approx 4.37$. In Fig.\ref{compare_fig} we compare the correlation length $\gamma\xi$ using the discrete and continuous-time formulation of the micromaser system for a typical setup of parameters. We observe the rapid convergence of the two formalisms already for a small value of $N$. In the sequel we will therefore make use of the fact that $\gamma\xi =1/\lambda_1 = 1/N\log(1/\kappa_1)$ if $N$ is large enough.

\vspace{1cm}
\bc{
\section{DETECTION EFFICIENCIES}
\label{sec:detection}}\ec

It has been emphasized in the literature that in the analysis of the micromaser system, when detection efficiencies has to be taken into account, one must distinguish between the occurrence of a certain detection event and the absence of such a detection event. This is so since the detection of an atom leaving the cavity necessarily gives rise to an instantaneous and non-local state reduction of the micromaser cavity radiation field. In Refs.\cite{walther94}  a non-linear equation of motion was derived taking such effects into account.
In studying the statistics of sequence events it was, however, shown in Ref.\cite{Herzog94} that one may make use of a linear equation of motion for non-normalized states. Recently it has been shown in detail how these different formalisms actually leads to the same physical results \cite{Johnson2001}. Below we follow the analysis of Refs.\cite{Herzog94,Johnson2001}   since we find it simpler to implement numerically.

Our calculation of joint or sequence probabilities when detection efficiencies must be taken into account follows the scheme outlined in Ref.\cite{Johnson2001}. There is simple prescription how to modify the analysis in Section \ref{sec:longcorr} to the situation with, in general, different detection efficiencies $\eta_+$ and $\eta_-$. The basic matrices $M(\pm)$ are naturally modified to ${\bar M(s)=\eta_s M(s)}$ with $s=\pm$ and we therfore also modify $M=M(+)+M(-)$ to ${\bar M } = \eta_+ M(+)+ \eta_- M(-)$. It is also convenient to define the matrix ${\bar M}_- = {\bar M} - M$. It then follows that the joint probability as given by Eq.(\ref{joint_k}) is modified according to
\begin{equation} \label{modified_joint_k}
    {\cal P}_k(s_{1},s_{2}) \rightarrow   {\bar {\cal P}}_k(s_{1},s_{2})= {\bar u}^{T}{\bar M}(s_2)
{\bar S}^{k}{\bar S}(s_1){\bar p}/{\cal N}~~~,
\end{equation}
where we have defined the matrix
\begin{equation} \label{modified_soperator}
    {\bar S}(s) = (1+L_{C}/N + {\bar M}_-)^{-1}{\bar M}(s)~~~.
\end{equation}
Here ${\bar S} = {\bar S}_+ + {\bar S}_-$ and ${\cal N}= {\eta}_+{\cal P}_+ + {\eta }_-{\cal P}_-$ is a normalization factor.  We also make the modification ${\cal P}(s) \rightarrow {\bar {\cal P}}(s)=\eta_s {\cal P}(s)/{\cal N}$.
It is of importance to observe that the stationary micromaser 
photon number distribution Eq.(\ref{p_n_eksakt})  still can be obtained from the stationary condition 
${\bar S}{\bar p}= {\bar p}$. The discussions in Section \ref{sec:longcorr} in the case of the discrete formulation of the micromaser system can now simply  be carried through by changing the relevant probabilities by the modified probabilities defined above and one, e.g., shows that $ {\bar {\cal P}}_k(s_{1},s_{2})= {\bar  {\cal P}}_k(s_{2},s_{1})$, which was shown explicitely for $k=0$ in Ref.\cite{Johnson2001}.  We therefore define the correlation function
\begin{eqnarray} \label{mod:gamma_D} 
  {\bar \gamma }_D(k) &=&
\frac{ {\bar {\cal P}} (+){\bar {\cal P}}(-) - {\bar {\cal P}}_k(+,-)}{ {\bar {\cal P}}(+) {\bar {\cal P}}(-)}\nonumber \\
&=& {\bar \gamma }_D e^{-k/R{ \bar \xi }_{D}} 
~,
\end{eqnarray} 
expressed in one of many equivalent manners as in Eq.(\ref{mod:gamma_D}). For $N$ large enough we can then write $\xi = { \bar \xi }_{D}$ if $\eta_+ = \eta_- =1$. Using a $1/N$ expansion similar to Eq.(\ref{eq:expansion}) we now find that
\begin{equation} \label{modified_expansion}
    {\bar S} = 1- \frac{1}{N}(L_{C}-N{\bar M}_-) + ({\bar M} - 1) = S~~~,
\end{equation}
apart from higher order terms in $(L_{C}-N{\bar M}_-)/N$ and higher orders in $1/N$. For sufficiently large $N$ we would then e.g. conclude that
\begin{equation} \label{modified_result_k}
    {\bar {\cal P}}_k(s_{1},s_{2})= {\bar u}^{T}{\bar M}(s_2)
e^{-\gamma Lt}{\bar S}(s_1){\bar p}/{\cal N}~~~.
\end{equation}
The  correlation length ${ \bar \xi }\equiv { \bar \xi }_{D} $ in Eq.(\ref{mod:gamma_D}) would then again be determined by the next-to-lowest eigenvalue of $L$ and, as a result, ${ \bar \xi }$ would actually not depend on the detection efficiencies at all and therefore equal to $\xi $. This was the claim as announced in Refs.\cite{Rekdal&Skagerstam&99a}. This conclusion is basically correct apart from a trivial correction which, unfortunately, was not taken into account. The physical explanation of this correction is simply  the fact that the time interval $t$ in Eq.(\ref{modified_result_k}) should be such that $Rt$ is average number of atoms which the observer  claims passes through the micromaser system, i.e. $Rt \simeq  k(\eta_+{\cal P}(+)+ \eta_-{\cal P}(-))$ and not $Rt \simeq k$. This means that the correlation length $\xi$ is "renormalized" to  ${\bar \xi} = (\eta_+{\cal P}(+)+ \eta_-{\cal P}(-))\xi$. A more formal proof of this assertion can be given as follows. Let us first consider the case of equal detection efficiencies, i.e. $\eta \equiv \eta_+ = \eta_-$. Since the correlation length ${\bar \xi}$ will be determined by the next-to-leading eigenvalue ${\bar \kappa}_1$ of the operator ${\bar S}$, i.e. $\gamma{\bar \xi}=1/N\log(1/{\bar \kappa}_1)$ for large $N$, we study the eigenvalue problem ${\bar S}{\bar p}_D = {\bar \kappa}{\bar p}_D$. As in Ref.\cite{ElmforsLS95} it is convenient to rewrite such an eigenvalue problem in an equivalent form, i.e. 
\begin{equation} \label{modified_eigen}
   \left( L_C - N(1+ \frac{\eta}{{\bar \kappa}} -\eta )(M-1))\right) {\bar p}_D = \eta N(\frac{1}{{\bar \kappa}} -1){\bar p}_D~~~.
\end{equation}
This equation can now be compared to the eigenvalue problem in the case of the continuous-time formulation of the micromaser system with $\eta_+ =\eta_- =1$, i.e.
\begin{equation} \label{cont_eigen}
   \left( L_C - N(M-1)\right) p_C = \lambda(N)p_C~~~,
\end{equation}
where we have made the $N$ dependence explicit in the eigenvalue $\lambda(N)$. By comparing Eqs.(\ref{modified_eigen}) and (\ref{cont_eigen}) we conclude that
\begin{equation} \label{detect_eigen}
   \lambda \left( N(1+ \frac{\eta}{{\bar \kappa}}-\eta ) \right) = \eta N(\frac{1}{{\bar \kappa}} -1) ~~~.
\end{equation}
\begin{figure}[htp]
\unitlength=0.5mm
\vspace{5mm}
\begin{picture}(160,140)(0,0)
\includegraphics{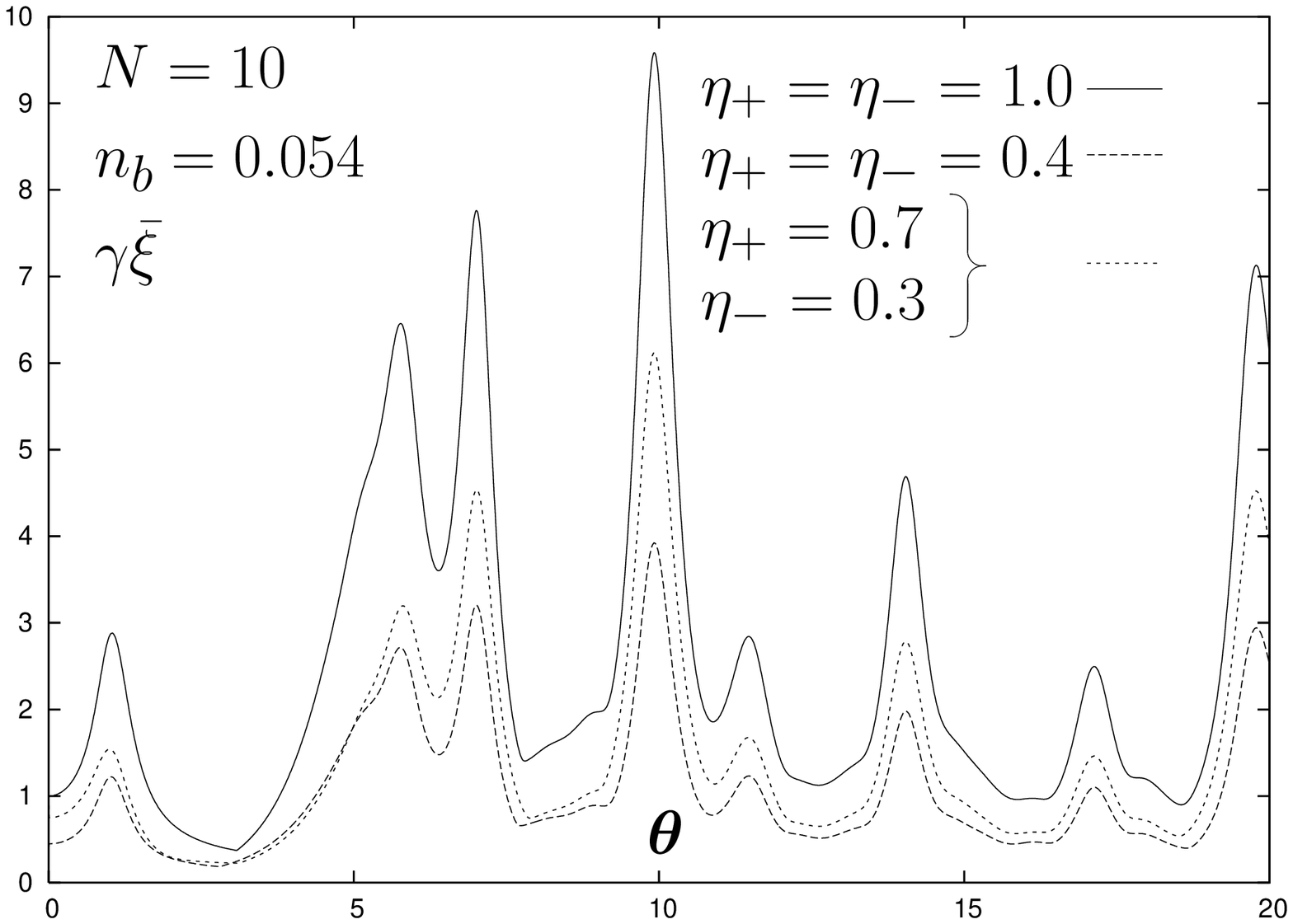}
 \end{picture}
\begin{picture}(160,140)(0,0)
\includegraphics{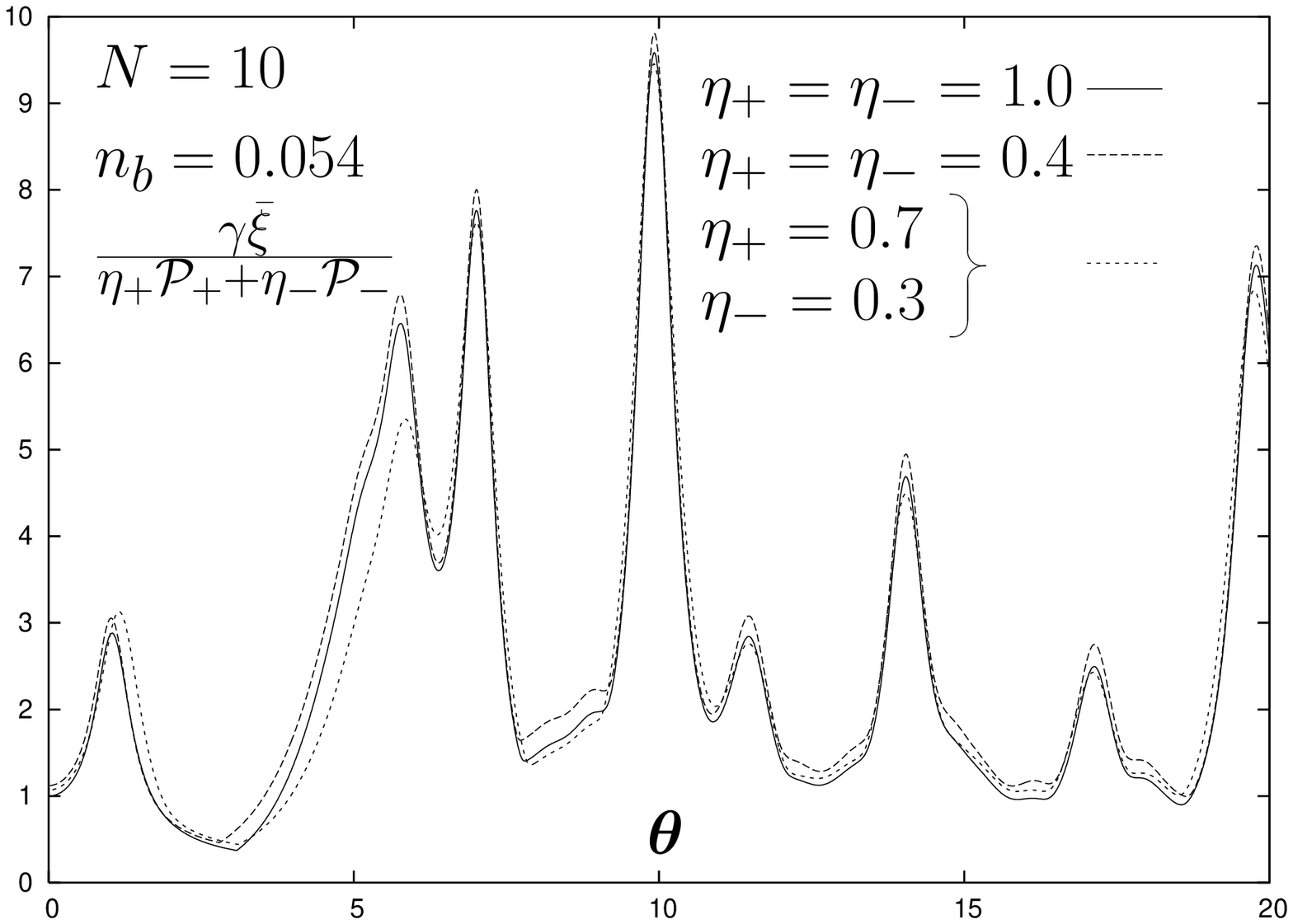}
 \end{picture}
\vspace{5mm}
\caption[]{\protect\small 
           In the upper graph the correlation length  $\gamma {\bar \xi}$ is given as a function of 
           $\theta$ for $N=10$ and $n_b=0.054$ for various values of detection efficiencies. With equal detection efficiencies $\eta = \eta_{+}=\eta_{-}$(lower curve) the correlation length is given by $\eta\gamma\xi $ with $\xi= {\bar \xi}(\eta_+=\eta_- =1)$. In the lower graph we consider the same parameters but a "renormalized" correlation $\gamma {\bar \xi}/(\eta_+{\cal P}_+ + \eta_-{\cal P}_-)$. }
\label{trapping_detect_figs}
\end{figure}

\noindent Since $\lambda (N)$ remains finite in the large-$N$ limit, we can write $ \lambda \left( N(1+ \eta/{\bar \kappa}-\eta ) \right) \rightarrow \lambda (N)$ for sufficiently large $N$ and we conclude that
\begin{equation} 
  \frac{1}{{\bar \kappa}_1} = 1 + \frac{\lambda_1}{\eta N} = 1+ \frac{1}{\eta\gamma\xi N}~~~.
\end{equation}
We therefore find that $\gamma {\bar \xi}= \eta \gamma \xi$ apart from $1/N$ corrections. A consequence of this analysis is that  $\eta N(1/{{\bar \kappa }}-1) M{\bar p}_D= \eta N(1/{{\bar \kappa }}-1){\bar p}_D$  in the large-$N$ limit as can be seen again by comparing Eqs.(\ref{modified_eigen}) and (\ref{cont_eigen}). This can be understood as follows. For large $N$  the components eigenvector ${\bar p}_D$ will be distributed around some large component. We can the replace the matrices $M(s)$ for $s=\pm$ by the diagonal matrices ${\cal P}(s)$, i.e. their mean-field values, in the expression $\eta N(1/{{\bar \kappa }}-1) M{\bar p}_D$ above. Extending the analysis above for a equal detection efficiencies to the case with different ones, we then find that $\gamma {\bar \xi}= (\eta_+ {\cal P}(+)+ \eta_-{\cal P}(-))\gamma \xi$ apart from $1/N$ corrections. Numerically, it turns out that the convergence of the correlation length as a function of $N$ is very rapid. In Figs. \ref{trapping_detect_figs} we study, as an example, the correlation length $\gamma {\bar \xi} \equiv \gamma {\bar \xi}_D$ for a moderate value of $N=10$ and other parameters adapted to  experimental data on trapping effects in the micromaser system \cite{W&V&H&W}. One verifies that $\gamma {\bar \xi}(\theta=0)=(N\log(1+1/N\eta_+))^{-1} $ in excellent agreement with numerical simulations in general. We find it remarkable that the scaling law for detection efficiencies of the ratio of correlation lengths ${\bar \xi}/\xi= \eta_+ {\cal P}(+)+ \eta_-{\cal P}(-) $ works so well for such a small value of $N$ as used in e.g. Figs.\ref{trapping_detect_figs}. Experimentally one could therefore measure the appropriate sequence probabilities, evaluate the correlation length and then renormalize the corresponding data with an easily calculable factor. The resulting correlation length so obtained is then the one that was predicted in Refs.\cite{ElmforsLS95}.

\vspace{1cm}
\bc{
\section{COLLECTIVE EFFECTS}
\label{sec:collective}}\ec

Collective effects are now due to the fact that during a time interval
$t$, such that $0 < t < \tau$, two atoms has jointly interacted with the
same cavity radiation field. An ideal and very special situation
corresponds $t=\tau$ which has
been discussed in great detail in Ref.\cite{wehner94} and also elsewhere \cite{two_atoms}. The 
general situation is more tedious   but straightforward to analyze and has been
discussed in great detail in Ref.\cite{haake97} in terms of an expansion in the
parameter $\epsilon = R\tau$. This parameter is supposed to be small, i.e. $\epsilon \ll 1$,  in order to be close to the
one-atom maser situation (see e.g. Appendix A in the second reference of Ref.\cite{ElmforsLS95}). We have reconsidered the analysis
of Ref.\cite{haake97}. The generator $L$ in the master equation 
Eq.(\ref{evolution_1}) is, up to first order terms in $\epsilon$,  modified according to
\begin{eqnarray}
\label{evolution_2}
 \frac{dp}{dt}= &-& \gamma \left( L_{C}-(1-2\epsilon )N(M-1)\right)p ~~~,
\nonumber \\ &+& \gamma N\frac{\epsilon}{\tau}\int_{0}^{\tau}dtu_2(t)p
~~,\nonumber \\
\end{eqnarray}
where the two-atom generator $u_2(t)$ describes two atoms that have jointly interacted with the cavity during the time interval $t$ such that $0< t < \tau$ and given explicitely in Ref.\cite{haake97}.  One finds that
the generator  $L$ in the master equation 
Eq.(\ref{evolution_1})
is replaced by $L_{tot} = L + L_{col}$, where $L_{col}$
describes the  two-atom collective effects with matrix
elements given by
\begin{eqnarray}
(L_{col})_{nm} = N\epsilon[(w_n(\tau)  + v_n(\tau))\delta_{n,m}
\nonumber \\ -
v_{n-1}(\tau)\delta_{n,m+1} -w_{n-2}(\tau)\delta_{n,m+2}]~~.\nonumber \\
\end{eqnarray}
Here we have defined
\begin{eqnarray}
\label{eq:vn}
v_n(\tau) &=& \frac{1}{\tau}\int_{0}^{\tau}dt  \left(\frac{}{}
[q_{n+1}(\tau-t)+b_n(t)][1- q_{n+1}(\tau-t) \right . \nonumber \\ &-& q_{n+2}(\tau-t)]
-q_{n+1}(\tau)[1-q_{n+1}(\tau)-q_{n+2}(\tau) ]\nonumber \\
  &-& \left. [c_n(t)+ d_n(t)][q_{n+1}(\tau-t)-q_{n+2}(\tau-t)]\frac{}{}\right)~~~,\nonumber \\
\end{eqnarray}
and
\begin{eqnarray}
\label{eq:wn}
w_n(\tau) = \frac{1}{\tau}\int_{0}^{\tau}dt \left(\frac{}{}
 c_n(t)+ q_{n+2}(\tau-t)q_{n+1}(\tau-t) \right. \nonumber \\ 
+ \left. b_{n}(t)q_{n+2}(\tau -t) + [c_n(t)+d_n(t)]q_{n+2}(\tau-t)\frac{}{}\right)
~~~.\nonumber \\
\end{eqnarray}
In these definitions we make use of the functions $q_n(t)\equiv q_n = \sin^2(gt\sqrt{n})$, 
\begin{eqnarray}
\label{def:bn}
  b_n(t)&=& \frac{n+1}{2n+3}q_{n+3/2}(2t)[1-q_{n+1}(\tau -t)]\nonumber \\
&-& \frac{1}{2}q_{n+3/2}(2t)q_{n+1}(\tau -t)~~~,
\end{eqnarray}
and
\begin{eqnarray}
\label{def:cn}
  c_n(t)&=&  \frac{n+1}{2(2n+3)}q_{n+3/2}(2t)[1-q_{n+1}(\tau -t)]\nonumber \\
&+& q_{n+3/2}^2(t)q_{n+1}(\tau -t)~~~,~~
\end{eqnarray}
and, finally, 
\begin{eqnarray}
\label{def:gn}
~d_n(t)&= &4\frac{n+1}{2n+3}\frac{n+2}{2n+3}
q_{n+3/2}^2(t)[1- q_{n+1}(\tau -t)] \nonumber \\
&+& \frac{n+2}{2(2n+3)}q_{n+3/2}(2t)q_{n+1}(\tau -t)~~~.~~~
\end{eqnarray}
\begin{figure}[htp]
\unitlength=0.5mm
\vspace{5mm}
\begin{picture}(160,140)(0,0)
\includegraphics{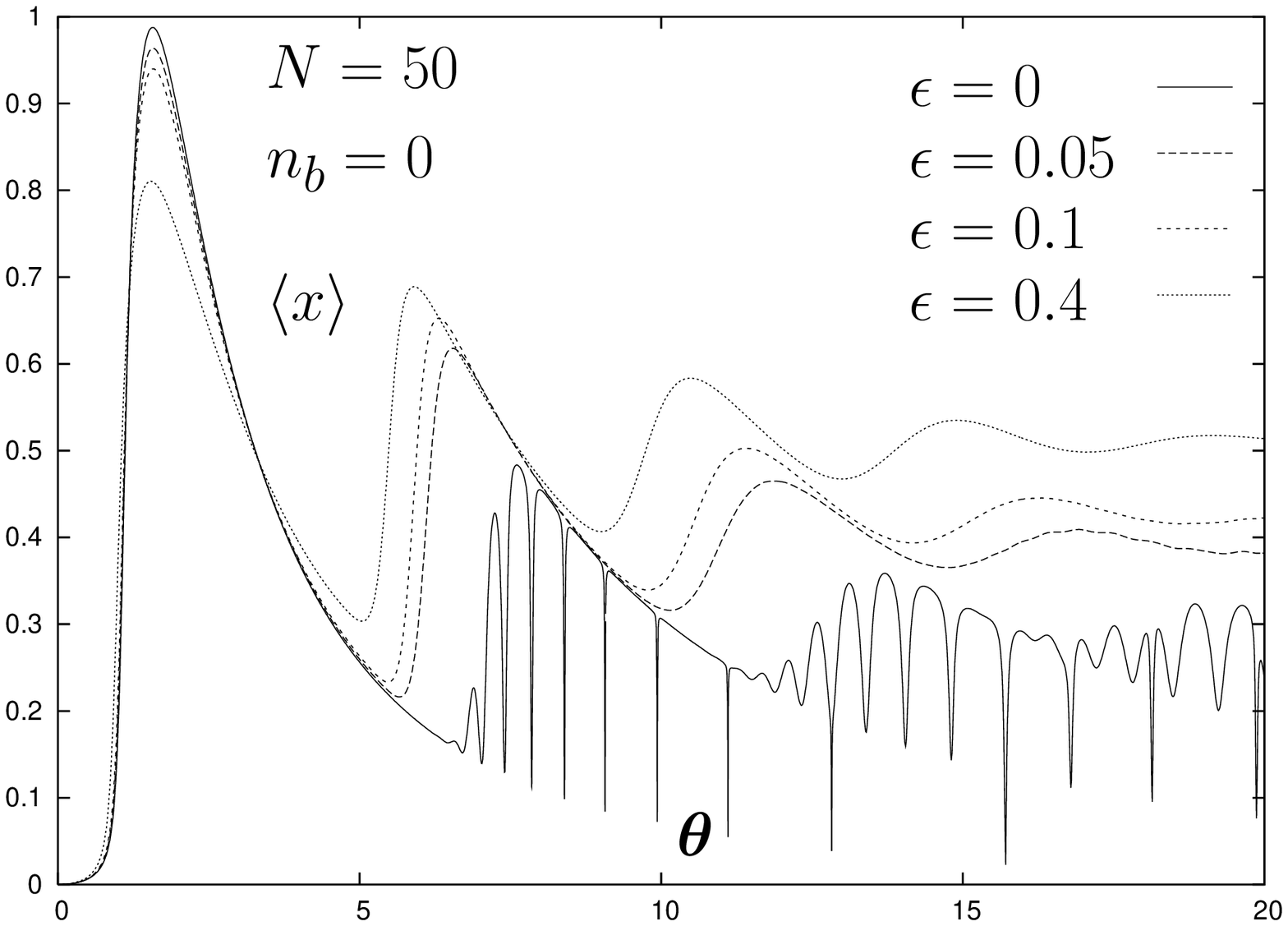}
 \end{picture}
\begin{picture}(160,140)(0,0)
\includegraphics{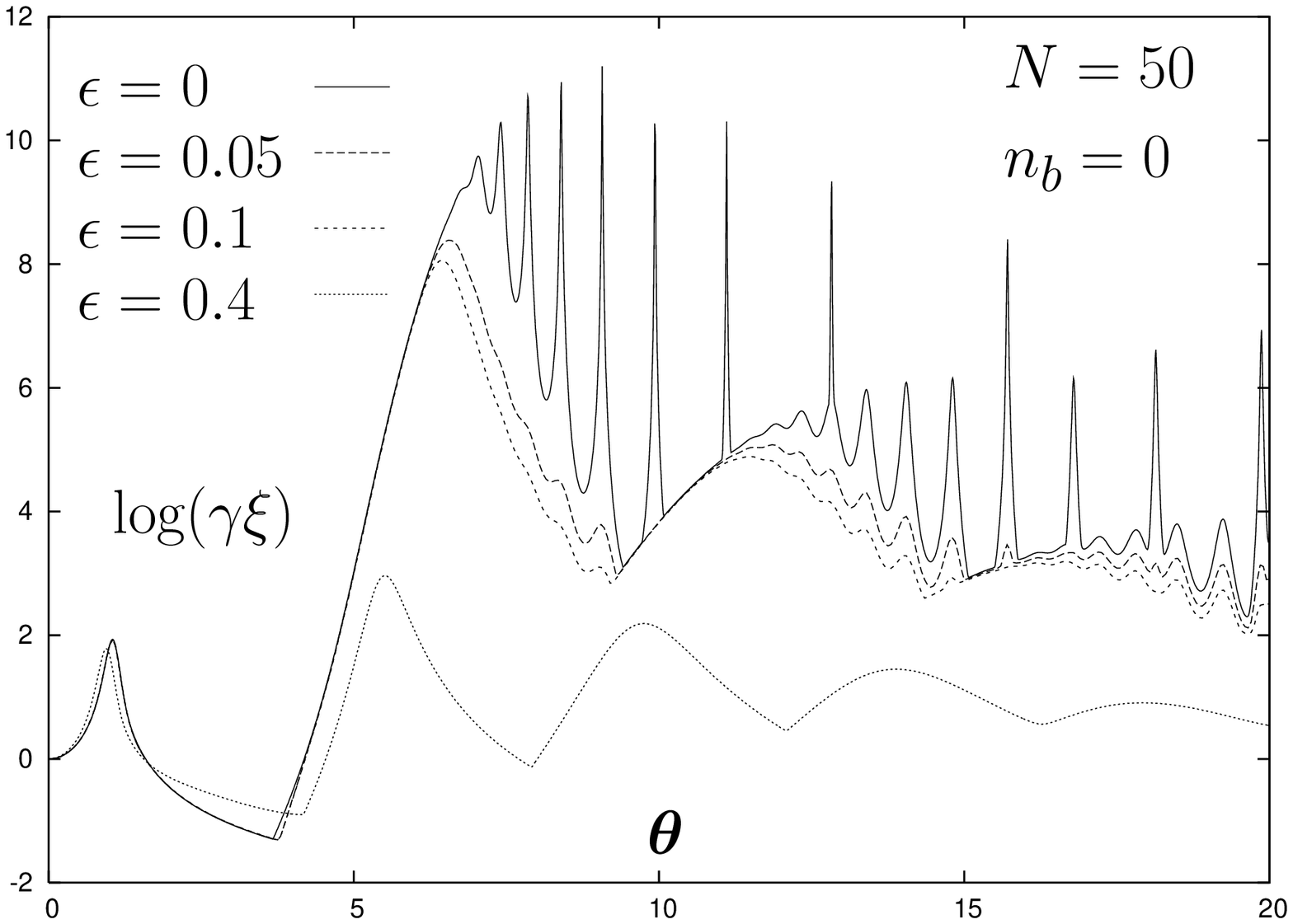}
 \end{picture}
\vspace{5mm}
\caption[]{\protect\small 
           In the upper graph the order parameter  $\langle x \rangle = \langle n/N \rangle$ is given as a function of 
           $\theta$ for $N=50$ and $n_b=0$ for various values of $\epsilon = R\tau$. It is only for $\epsilon =0$ that trapping-state peaks are visible. In the lower graph the correlation length $\gamma\xi$ is given for the same set of parameters. Even though the trapping-state peaks vanishes as   $\epsilon $ increases, the correlation length can still be large for $\epsilon \neq 0$. The peaks at $\theta \simeq 1$ correspond to the first maser transition.}
\label{Kolobov_figs}
\end{figure}
\noindent Our final expression for $L_{tot}$ is, in fact, 
in agreement with the result of Ref.\cite{haake97} using a slightly different
  notation. It is rather straightforward to implement the matrix elements of $L_{tot}$ in a numerical routine
in order to find the new stationary distribution ${\bar p}$ corresponding to the eigenvalue $\lambda_0=0$ of $L_{tot}$ to be used in evaluating various expectation values. The correlation length is again determined by the next-to-lowest eigenvalue $\lambda_1$ of $L_{tot}$. In the numerical work it turns out to be sufficient to use 200x200 matrices for both $L$ and $L_{tot}$ in order to obtain the accuracy of the graphs as presented in the present paper. In Fig.\ref{Kolobov_figs} we show the results of a numerical evaluation of the order parameter $\langle x \rangle$ and the correlation length $\gamma\xi$ for $N=50$ in a vacuum configuration ($n_b=0$). 

\begin{figure}[htp]
\unitlength=0.5mm
\vspace{5mm}
\begin{picture}(160,140)(0,0)
\includegraphics{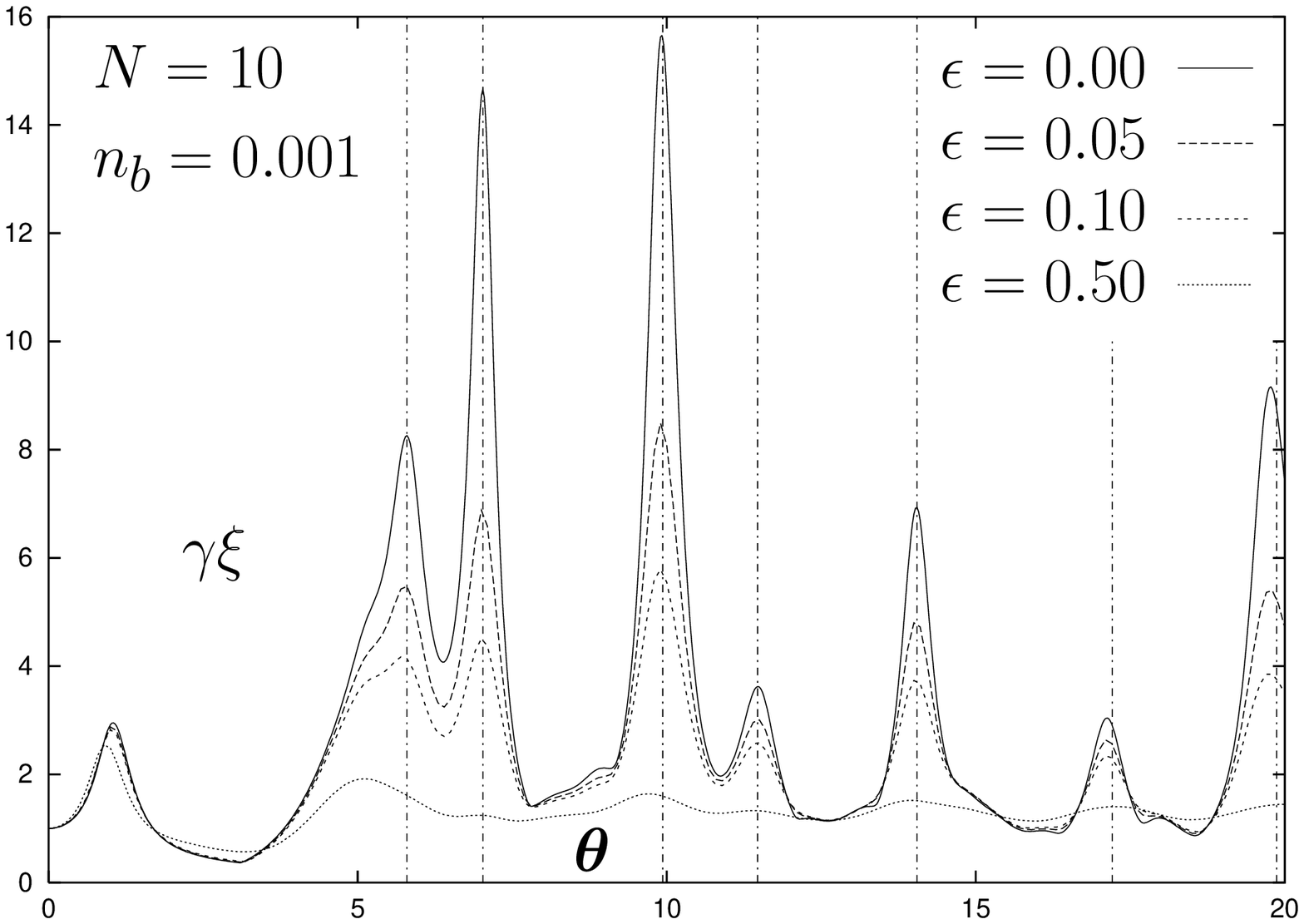}
 \end{picture}
\vspace{5mm}
\caption[]{\protect\small 
           The correlation length  $\gamma \xi$ as a function of 
           $\theta$ for various values of $\epsilon$.
           The vertical lines indicate the trappings values of $\theta =\pi \sqrt{N}(1/\sqrt{3}, 1/\sqrt{2},1,2/\sqrt{3},\sqrt{2}, \sqrt{3},2) $.  }
\label{corr_fig_001}
\end{figure}

\begin{figure}[htp]
\unitlength=0.5mm
\vspace{5mm}
\begin{picture}(160,140)(0,0)
\includegraphics{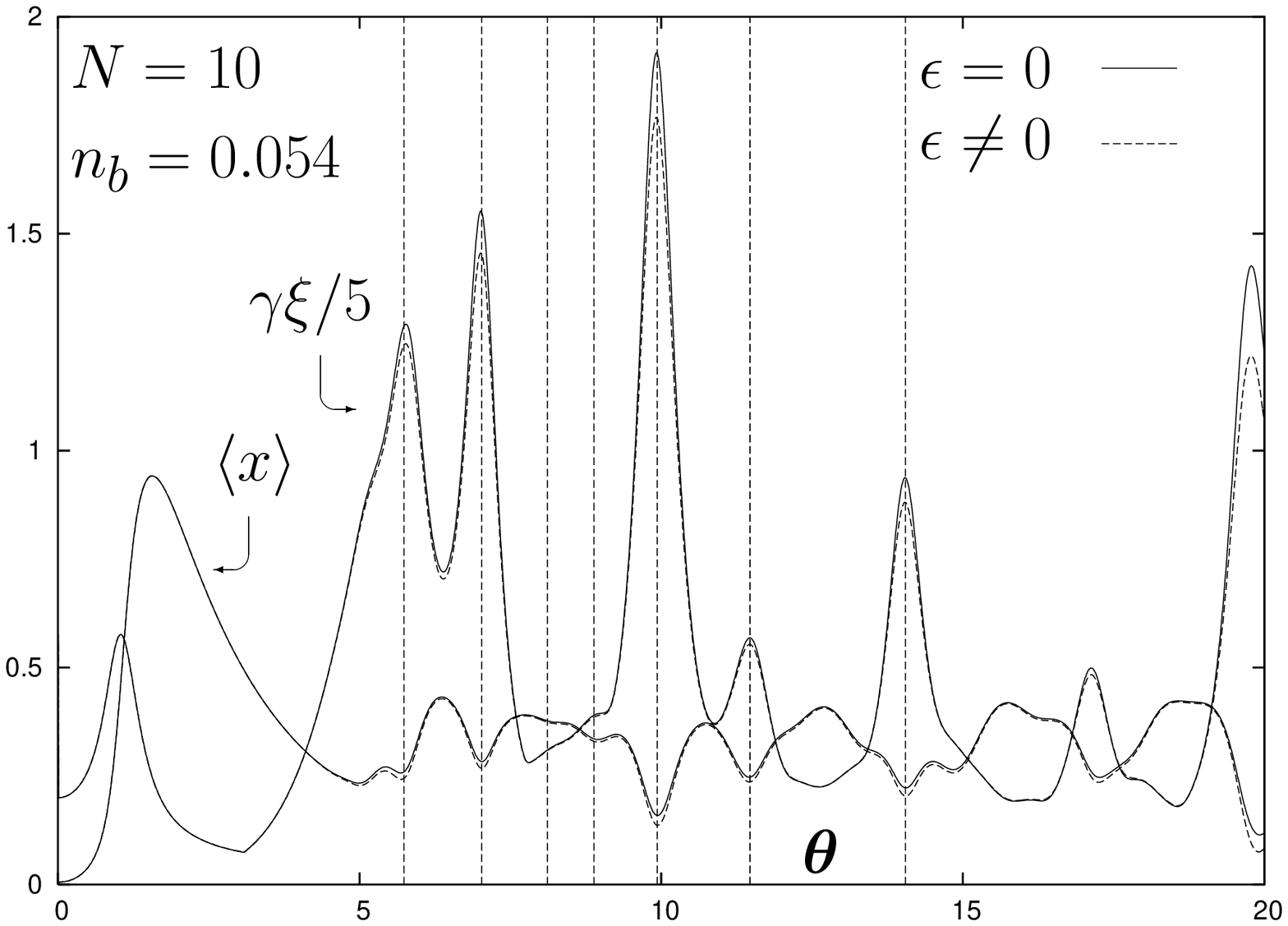}
 \end{picture}
\vspace{5mm}
\caption[]{\protect\small 
           The order parameter $\langle x \rangle$ and the correlation length  $\gamma \xi/5$ as a function of 
           $\theta$ with no collective effects taken into account ($\epsilon =0$) or $\epsilon =R\tau = \theta\sqrt{N}\gamma/g$. The Rabi frequency is as in the experimental study of trapping-states in Ref.\cite{W&V&H&W}, i.e. $g=39~kHz$ and $\gamma = 10~s^{-1}$.
           The vertical lines indicate the experimentally observed trappings values of $\theta =\pi \sqrt{N}(1/\sqrt{3}, 1/\sqrt{2},2/\sqrt{6},2/\sqrt{5},1,2/\sqrt{3},\sqrt{2}) $ as reported in Ref.\cite{W&V&H&W}. }
\label{garching_trapping_99}
\end{figure}

\noindent Our results basically agree with the corresponding numerical results of Ref.\cite{haake97} even though we appearently  have a higher numerical precision. As argued in Ref.\cite{haake97}, effects of trapping states at $\theta =k\pi \sqrt{N/n}$, for $k,n=1,2,...$, vanishes as $\epsilon \neq 0$. This is, however, not so for the correlation length $\gamma\xi$ as indicated in the lower figure of  Fig.\ref{Kolobov_figs}. In fact, if we consider even a lower value of $N=10$ and $n_b=0.001$ as in Fig.\ref{corr_fig_001}, trapping-state effects are  clearly seen in the correlation length. In Fig.\ref{corr_fig_001} the vertical lines corresponds to the clearly visible trapping states with $\epsilon = 0$. In the experimental study of trapping-states in Ref.\cite{W&V&H&W} the parameters are, however, varied in such a way that the parameter $\epsilon$ is not constant but is given by $\epsilon = R\tau = \theta\sqrt{N}\gamma/g$ as in Fig.\ref{garching_trapping_99}, where we have chosen the cavity temperature to be $T=0.3 K$, corresponding to $n_b = 0.054$. The average photon lifetime in the cavity corresponding to Fig.\ref{garching_trapping_99} is 0.1 $s$, i.e. $\gamma = 10~s^{-1}$. The vertical lines in Fig.\ref{garching_trapping_99} correspond to the trapping-states as considered in Ref.\cite{W&V&H&W} and studied in terms of the atomic inversion.   With detection efficiencies taken into account, the appropriate definition of the atom inversion $I(\tau)$ is given by 
\begin{eqnarray}
\label{def:I}
I(\tau) &=& \frac{\eta_{+}{\cal P}(+)-\eta_{-}{\cal P}(-)}{\cal N} \nonumber \\  &=& \frac{(\eta_{+}+\eta_{-}) }{\cal N}\frac{{\displaystyle n_b}}{{\displaystyle N}} - \frac{\eta_{+}+\eta_{-}}{\cal N}\langle x \rangle ~~~,
\end{eqnarray}
where is ${\cal N}$ is the normalization factor $\eta_{+}{\cal P}(+)+\eta_{-}{\cal P}(-)$. Eq.(\ref{def:I}) gives the general relation between the atomic inversion $I(\tau)$ and the order parameter $\langle x \rangle$. We observe that if $\eta \equiv \eta_{+}=\eta_{-}$ then $I(\tau)$ is independent of $\eta$. It is clear form our Fig.\ref{garching_trapping_99} that trapping effects are much more visible in the correlation length than in the order parameter $\langle x \rangle$. Its is also clear that collective effects are small with the set of micromaser parameters chosen. Instead detection efficiency will be a major correction to the theoretical values.
In Section \ref {sec:detection} we have seen how the correction length is to be corrected for due to detection efficiencies. In view of these results it therefore appears that trapping-states can be more clearly revealed experimentally in terms of  the correlation length rather than the atomic inversions or, equivalently, the order parameter $\langle x \rangle$.

\vspace{0.5cm}
\bc{
\section{FINAL REMARKS}
\label{sec:final}
}\ec
\vspace{0.5cm}

In conclusion, we have studied how to take detection efficiencies into account when comparing experimental data for the correlation length to the theoretical predictions. We have found a remarkable and simple scaling relation that connects observational data and the theoretical prediction. We have also studied two-atom collective effects in terms of the natural parameter $\epsilon = R\tau$. Even though cumbersome, the calculations are in principle straightforward. Trapping effects appear to suppressed with increasing values of $\epsilon$. We have, however, seen that in a realistic experimental situation, as discussed in e.g. Ref.\cite{W&V&H&W}, collective effects are, nevertheless, small and detection efficiencies are more important to take into account. As we have shown, detection efficiencies can, however, be taken into account in a straightforward manner.

   With increasing values of $N$, signals due to the micromaser phase transitions 
   become more pronounced. General methods, which are exact in the large-$N$ limit, for computing characteristic features of these phase transition have been presented in Refs.\cite{ElmforsLS95,Rekdal&Skagerstam&99a}. Detection efficiencies will only imply a calculable rescaling as discussed in Section \ref{sec:detection}.

\begin{figure}[htp]
\unitlength=0.5mm
\vspace{5mm}
\begin{picture}(160,140)(0,0)
\includegraphics{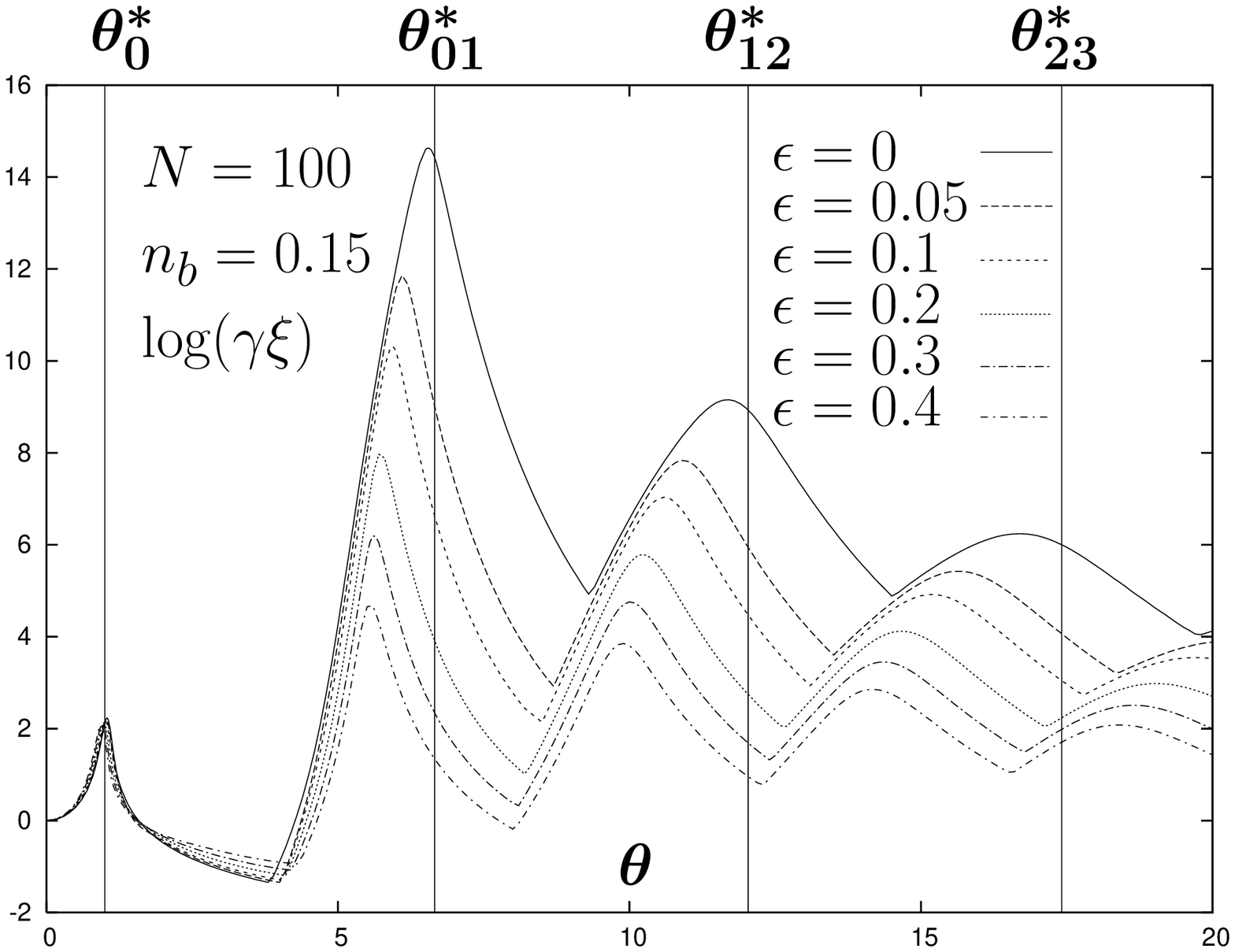}
 \end{picture}
\vspace{5mm}
\caption[]{\protect\small 
         The correlation length  $\gamma \xi$ as a function of 
           $\theta$ for various values of $\epsilon$. 
           The vertical lines indicate the large $N$ values for the maser phase transitions at $\theta = \theta^{*}_{0}=1 , \theta^{*}_{01}\approx 6.6610, \theta^{*}_{12}\approx 12.035, \theta^{*}_{23}\approx 17.413 $. With $\langle x\rangle$ as an order parameter, the transition at $\theta= \theta^{*}_{0}$  is second order while the other are first order transitions. }
\label{Phase_Transitions}
\end{figure}

Due to the random arrival
statistics of the pump atoms there is, however, in an actual experimental situation
a finite probability of more than
one pump atom in the cavity \cite{wehner94}. Since the average number of pump atoms 
inside the cavity is $\epsilon \equiv \tau R =\gamma\sqrt{N}\theta/g$, this parameter, as we have seen above,
naturally parameterize the probability of collective
pump atom effects. If $\epsilon$ is assumed to be constant and small, i.e. $\epsilon \ll 1$, we see that
the dimensionless pump parameter $N$ is bounded by
$\sqrt{N} \ll g/\gamma\theta$. Arbitrarily large values of $N$ can then only be reached by
making $g/\gamma$ arbitrarily large which, of course, is difficult to achieve in a real
experimental realization of the micromaser. Alternative $\epsilon = \theta \sqrt{N}g/\gamma$ should be small as $\theta$ varies.  As we have seen in the present paper, if $\epsilon$ is sufficiently small,  
corrections to the observables as discussed in the present paper can 
be calculated in a rather straightforward manner.

At finite $N$ and including collective pump atom effects, signals of the large
$N$ phase transitions in the order parameter
 $\langle x\rangle $ are still clearly exhibited, at least in the case 
when the pump atoms are prepared in the excited state and at resonance with cavity radiation field.  The critical point $\theta^{*}_{0}$ of the first
second-order maser transition remains the same with $\langle x \rangle$ used as a natural order parameter. As seen from Eq.(\ref {evolution_2}) when taking collective effects into account for a sufficiently small $\epsilon$, one-atom effects are modified by a renormalization $N \rightarrow \exp(-2\epsilon)N$, where $\exp(-2\epsilon)\approx 1-2\epsilon$ is the probability for one-atom events in the micromaser cavity. We therefore expect that the critical parameters $\theta^{*}_{kk+1} \propto \sqrt{N}$ of first-order transitions
are changed to ${\bar\theta}^{*}_{kk+1}\equiv \exp(-\epsilon)\theta^{*}_{kk+1}$. By making use of  the general results of Refs.\cite{ElmforsLS95,Rekdal&Skagerstam&99a}  concerning the $N$ dependence of  the peak values of the correlation length these are changed accordingly, and we find that
$\log(\gamma\xi)_{crit}$ is changed to 
$\log(\gamma{\bar \xi})_{crit}=\exp(-\epsilon)\log(\gamma\xi(\theta = {\bar\theta}^{*}_{kk+1}))$. In Fig.\ref{Phase_Transitions} we illustrate these effects in the case $N=100$ and $n_b=0.15$. For small values of $\epsilon$ we find that the scaling behavior given above for ${\bar\theta}^{*}_{kk+1}$ and $\log(\gamma{\bar \xi})_{crit}$  are  indeed compatible with the exact results. 

As in Ref.\cite{haake97} we also find that our perturbative methods gives meaningful results even when the expansion parameter $\epsilon $ is  large. In the present paper we have considered a limited range of micromaser parameters. If atoms and the radiation field are not at resonance one can e.g. extend the results of the  present paper by making use of the methods of Ref.\cite{Rekdal&Skagerstam&99a} and the procedures outlined in the present.
%
{\begin{center}
{ \bf ACKNOWLEDGMENT }
\end{center}}
%
   The author wishes to thank B. T. H. Varcoe and H. Walther and for discussions
  and H. Walther for providing a
   guide to the progress in experimental work over the years. G\"{o}ran Wendin  is acknowledged
  for hospitality during the completion of this work. NorFA is
  acknowledged for its support.

\end{document}